\documentclass[authoryear,table]{elsarticle}

\usepackage[english]{babel}
\usepackage{fancyhdr} 
\usepackage{url} 
\usepackage{caption} 
\usepackage{subcaption} 
\usepackage{natbib} 
\usepackage{multirow} 
\usepackage{float} 
\usepackage{amssymb}
\usepackage[intlimits]{amsmath} 
\usepackage{amsthm} 
\usepackage[german]{nomencl}
\usepackage{booktabs} 
\usepackage{geometry}
\usepackage{enumitem}
\usepackage{tikz}
\usepackage{xcolor}
\usepackage{epstopdf}
\usepackage{rotating}
\usepackage{hhline}
\usepackage{soul}
\usetikzlibrary{calc,positioning,shapes.multipart,decorations.pathreplacing,shapes.arrows}

\definecolor{red1}{RGB}{240,80,80}
\definecolor{red2}{RGB}{246,136,93}
\definecolor{yellow1}{RGB}{247,125,47}
\definecolor{yellow2}{RGB}{255,152,96}
\definecolor{green1}{RGB}{210,247,123}
\definecolor{green2}{RGB}{130,194,91}
\definecolor{grey1}{RGB}{211,211,211}

\def\TextShift{15pt}

\tikzset{
  myrect/.style={   
  rectangle,
	minimum width=#1*20,
	minimum height=0.8cm,
   draw,
    anchor=west,
  },
  mytext/.style={
    rectangle,
    draw=#1!70!black,
    fill=#1,
    align=center,
    line width=1pt,
    font=\sffamily
  },
  mytextb/.style={
    mytext=#1,
    anchor=north
  },
  mytexta/.style={
    mytext=#1,
    anchor=south
  }
}

\newcommand\AddText[5][]{
  \if#5l\relax
    \node[mytextb=#2,yshift=-\TextShift,#1] 
      at (part#4.south west) {\strut#3\strut};
  \fi
  \if#5L\relax
    \node[mytexta=#2,yshift=\TextShift,#1] 
      at (part#4.north west) {\strut#3\strut};
  \fi
  \if#5m\relax
    \node[mytextb=#2,yshift=-\TextShift,#1] 
      at ( $ (part#4.south west)!0.5!(part#4.south east) $ ) {\strut#3\strut};
  \fi
  \if#5M\relax
    \node[mytexta=#2,font=\bf ,yshift=\TextShift,#1] 
      at ( $ (part#4.north west)!0.5!(part#4.north east) $ ) {\strut#3\strut};
  \fi
  \if#5r\relax
    \node[mytextb=#2,yshift=-\TextShift,#1] 
      at (part#4.south east) {\strut#3\strut};
  \fi
  \if#5R\relax
    \node[mytexta=#2,yshift=\TextShift,#1] 
      at (part#4.north east) {\strut#3\strut};
  \fi
}

\newcommand\TimeLine[1]{%
\coordinate (part0); 
\foreach \Longitud/\Color/\Texto [count=\ti] in {#1}
{
  \node[
    myrect=\Longitud,
    fill=\Color,
    right=of part\the\numexpr\ti-1\relax
    ] 
      (part\ti)
      {};
  \node[font=\normalsize]
    at (part\ti.center) {\Texto};  
  \gdef\lastpart{\ti}
}
}

\newcommand\TimeLinetwo[1]{%
\coordinate[below=1cm of part0] (part0); 
\foreach \Longitud/\Color/\Texto [count=\ti] in {#1}
{
  \node[
    myrect=\Longitud,
    fill=\Color,
    right=of part\the\numexpr\ti-1\relax
    ] 
      (part\ti)
      {};
  \node[font=\normalsize]
    at (part\ti.center) {\Texto};  
  \gdef\lastpart{\ti}
}
}

\newcommand\TimeLinethree[1]{%
\coordinate[below=1cm of part0] (part0); 
\foreach \Longitud/\Color/\Texto [count=\ti] in {#1}
{
  \node[
    myrect=\Longitud,
    fill=\Color,
    right=of part\the\numexpr\ti-1\relax
    ] 
      (part\ti)
      {};
  \node[font=\normalsize]
    at (part\ti.center) {\Texto};  
  \gdef\lastpart{\ti}
}
}

\newcommand\TimeLinefour[1]{%
\coordinate[below=1cm of part0] (part0); 
\foreach \Longitud/\Color/\Texto [count=\ti] in {#1}
{
  \node[
    myrect=\Longitud,
    fill=\Color,
    right=of part\the\numexpr\ti-1\relax
    ] 
      (part\ti)
      {};
  \node[font=\normalsize]
    at (part\ti.center) {\Texto};  
  \gdef\lastpart{\ti}
}
}

\newcommand\TimeLineinvis[1]{%
\coordinate[below=1cm of part0] (part0); 
\foreach \Longitud/\Color/\Texto [count=\ti] in {#1}
{
  \node[
    myrect=\Longitud,
    white,
    right=of part\the\numexpr\ti-1\relax
    ] 
      (part\ti)
      {};
  \node[font=\footnotesize]
    at (part\ti.center) {\Texto};  
  \gdef\lastpart{\ti}
}
}

\newcommand\mydaylength{0.55}

\makenomenclature
\usepackage[pdfborder={000},colorlinks=true,linkcolor=blue,citecolor=blue]{hyperref}
\usepackage{listings}
\makeindex

\usepackage{color}

\definecolor{white}{rgb}{1,1,1}
\definecolor{tikcol1}{rgb}{1,.8,.8}
\definecolor{tikcol2}{rgb}{0.7,1,.7}
\definecolor{tikcol3}{rgb}{0.5,.8,1}
\definecolor{tikcol4}{rgb}{0.8,.8,.8}

\definecolor{b}{rgb}{0,0,.8}	
\definecolor{g}{rgb}{0,.6,0}	
\definecolor{n}{rgb}{0,0,0}	
\definecolor{h}{rgb}{0.4,0.2,0.2}	
\definecolor{v}{rgb}{0.2,0.6,0}

\newcommand{\E}{{\mathbb E}}

\newcommand{\G}{{\mathbb G}}

\newcommand{\R}{{\mathbb R}}
\newcommand{\T}{{\mathbb T}}

\newcommand{\X}{{\mathbb X}}

\newcommand{\bsB}{\boldsymbol B}

\newcommand{\bsY}{\boldsymbol Y}

\newcommand{\bsone}{\boldsymbol 1}

\newcommand{\bsbeta}{\boldsymbol \beta}

\newcommand{\bseps}{\boldsymbol \varepsilon}

\newcommand{\bsgamma}{\boldsymbol \gamma}

\newcommand{\eps}{{\varepsilon}}

\DeclareMathOperator*{\argmin}{arg\,min}

\newcommand{\ov}\overline
\newcommand{\what}{\widehat}
\newcommand{\wtilde}{\widetilde}

\newcommand{\rig}\right
\newcommand{\lef}\left
\newcommand{\nf}\normalfont

\newcommand{\MAE}{\text{MAE}} 
\newcommand{\MMAE}{\text{MMAE}}

\definecolor{rickgreen}{rgb}{0,0.6,0}

 \definecolor{strikeout}{rgb}{0.5,0.5,.5}
\definecolor{newchange}{rgb}{0.8,0,0}
 \newcommand{\OLD}[1]{}
 \newcommand{\OLDg}[1]{}
 \newcommand{\NEW}[1]{\color{black} #1 \color{black}}

\begin{document}

\title{\NEW{Short- to} Mid-term \NEW{Day-Ahead} Electricity Price Forecasting Using \NEW{Futures} \OLD{Future Data}}

\author{Rick~Steinert}
\ead{steinert@europa-uni.de}
\address{Europa-Universit\"at Viadrina, Fakult\"at f\"ur Wirtschaftswissenschaften, Gro\ss e Scharrnstra\ss e 59, 15230 Frankfurt (Oder), Germany}

\author{Florian~Ziel}
\ead{Florian.Ziel@uni-due.de}
\address{Universit\"at Duisburg-Essen, Fakult\"at f\"ur Wirtschaftswissenschaften, Berliner Platz 6-8, 45127 Essen}

\begin{frontmatter}
\begin{abstract}
Due to the liberalization of markets, the change in the energy mix and the surrounding energy laws, electricity research is a dynamically altering field with steadily changing challenges. One challenge \NEW{especially for investment decisions} is to provide reliable \NEW{short to} mid-term forecasts despite high variation in the time series of electricity prices. This paper tackles this issue in a promising and novel approach. By \OLD{utilizing} \NEW{combining} the \OLD{high} precision of econometric autoregressive models \NEW{in the short-run with} \OLD{and} the expectations of market participants reflected in future prices \NEW{for the short- and mid-run} we show that the forecasting performance can be vastly increased while maintaining hourly precision. We investigate the day-ahead electricity price of the EPEX Spot for Germany and Austria and setup a model which incorporates the Phelix future of the EEX for Germany and Austria. The model can be considered as an AR24-X model with one distinct model for each hour of the day. We are able to show that future data contains relevant price information for future time periods of the day-ahead electricity price. \NEW{We show that relying only on deterministic external regressors can provide stability for forecast horizons of multiple weeks.}  By implementing a fast and efficient lasso estimation approach we demonstrate that our model can outperform several other models in the literature.
\end{abstract}
\end{frontmatter}

\section{Introduction} \label{Introduction}
Modeling and forecasting electricity prices have become an important and broad part of economic research during the last decades. The specifics of electricity prices, also known as stylized facts as well as the \NEW{due to new laws} rapidly changing market conditions especially in Europe and Germany have promoted this development. Moreover, the data transparency has tremendously increased during the last years, either by law or by negotiated agreement data for e.g. electricity consumption, production, prices and even planned capacities can be downloaded via different sources like ENTSO-E or the exchanges themselves. The electricity exchanges also expanded their product portfolio by launching new electricity related products like new block products, derivatives or complete new spot auctions as e.g. the EXAA GreenPower auction. Even though these changes will provide the informed decision maker with more valid options, it also increases the complexity of the decision making process. 

One of the research approaches which tries to incorporate the changing market conditions is the econometric perspective, which in general constructs models which aim to capture the underlying behavior of the electricity price time series and can provide forecasts afterwards. These forecasts can help market participants in their decision making for e.g. investment decisions. \NEW{Moreover, forecasts offer different utility dependent on their forecasting horizon. Forecasts of only a few days in advance can help electricity companies to adjust their production planning. For instance, if an owner of a pumped-storage hydroelectricity plant has information on extremely low prices in the future they can easily schedule their generation of electricity by releasing their water reservoir now and refill it later when the electricity price is low. Medium- or long-term forecasts can help market participants to identify investment opportunities in the long-run, e.g. when the decision of the construction of a new wind power plant is considered, as they need to have reliable information on future cash-flows for their product. This is especially important in Germany, as the market premium which producers of renewable energy receive is calculated according to the attachment 1 of \S 23a EEG (``Erneuerbare-Energien-Gesetz") by using, among others, the average monthly spot prices of the EPEX SE.}

Econometric models usually use the intertemporal correlation structure of day-ahead electricity prices and combine them with external fundamental or stylized facts related regressors to provide good forecasts\NEW{, see for instance \cite{weron2014electricity} for an extensive review of different models}. However, these models usually struggle when it comes to mid- or even long-term horizon forecasting. The reason for that is mainly, that every non-deterministic regressor like electricity load, wind- and solar power production, water reservoir levels or fuel prices have to be forecasted as well. This means that the forecaster has not only the task to come up with a good model for the electricity price but also for the regressors, even though both time series may come from very different disciplines of research. Moreover, due to their autoregressive structure every error in forecasting of one of the series will have an impact on any consecutive forecasting time point, depending on the magnitude of the intertemporal correlation of the time series itself and especially the residuals. \NEW{Some authors therefore try to either use already forecasted regressors or only lags of external regressors (e.g. \cite{bunn2016analysis} or \cite{hagfors2016using}). This in turn leads to a situation where the forecasting horizon is restricted to the lowest used lag of the external regressors.} When the day-ahead price of electricity is concerned, this means that due to the usual hourly resolution, forecasting e.g. four weeks leads to $4\times7\times24=672$ points in time which have to be forecasted. \NEW{Simple autoregressive models also converge quickly to their mean, which make them incapable of forecasting longer forecasting horizons}\citep{keles2012comparison}.

Therefore, we want to setup a model, which is capable of generating reliable \NEW{short to} mid-term forecasts of up to four weeks by using regressors, which provide a preferably long deterministic structure, which means that we do not have to forecast them for as long as possible. For this we have decided to use the EPEX day-ahead electricity spot price of Germany and Austria and combine it with the EEX Phelix futures which have a cash settlement based on the \NEW{average} EPEX spot price for different time horizons. \OLD{We will use the term mid-term to refer to a horizon ranging from one month to one year, even though the literature is not consentaneous on this.} \NEW{As the literature is not consentaneous on the distinction between short-term, mid-term and long-term, we decided to declare the forecasting horizon we use as short- to mid-term. Our forecasting horizon comprises 1 to 28 days. In the next paragraphs, whenever we list a paper according to their forecasting horizon, we follow their own definition of the term.}

The literature on mid- to long-term electricity price forecasting is very scarce \citep{yan2013mid}. This holds especially true for econometric modeling. \cite{maciejowska2016short} for instance utilize an autoregressive modeling approach to forecast the UK electricity price for up to 45 days. The authors compare the difference in forecasting accuracy of, among others, AR-models with hourly precision and AR-models which only use the daily average. They find that in the mid-term simpler models without hourly resolution seem to be superior against more complex models which keep the complex hourly structure, while in the short-term this the relation is the other way round. Moreover, they also find that including regressors did not always lead to better forecasts, the inclusion of $\text{CO}_2$ prices for instance weakened the accuracy in general due to problems with forecasting this time series. 

In the study of \cite{ziel2017electricity} the authors apply an econometric autoregressive approach towards the sale and purchase curves of the EPEX day-ahead electricity price. By a simulation study they can replicate the market situation and provide mid- to long-term probabilistic forecasts for the \NEW{electricity price as well as all other related components}\OLD{complete market situation}. For their study they use the auction bids as well as external regressors like wind and solar power. By evaluating coverage probabilities they are able to compare their probabilistic forecasting values with the real electricity price time series and state that given the long horizon the models tends to have promising results.

Other approaches for mid- and long-term forecasting originate from other fields of electricity price research, e.g. heuristics as in \cite{yan2015midterm} or fundamental models like in 
\cite{bello2016probabilistic, bello2017medium}.

Nevertheless the relationship between spot and future products is an extensive field of research in finance and in energy economics as well. However, the typical relationship of futures can be described by the difference in expectation about spot prices and the price of a future, which in commodities research is due to the participants  \NEW{necessity of getting a} premium for \NEW{storing} \OLD{holding} a specific asset\OLD{ rather than buying the future product }\citep{weron2014revisiting}. But as electricity prices cannot be stored easily, this relationship tends to be more complex. The basic relationship is typically described as follows: (see e.g. \cite{benth2008pricing})
\begin{equation} \label{eq_Risk_Premium}
\text{RP}_{t,\T} = \E[P_{\T}|I_t] - F_{t,\T}, 
\end{equation}
where $\E[P_{\T}|I_t]$ is the expected electricity price of delivery period $\T$ based on the information set $I_t$ at time $t<T$ for $T\in \T$. $\text{RP}_{t,\T}$ represents the before mentioned risk premium and $F_{t,\T}$ the price of the future in $t$ for period $\T$ with the electricity price as underlying. 
Usually the delivery period is an interval $\T = [T_1, T_2]$ with $T_1<T_2$. 
However, in practice futures are often quoted with a corresponding maturity. For instance, the Phelix Day Base Future with maturity 2 refers to a delivery period of all 24 hours of the day starting with the first hour of the day 2 days after the product was traded.
This is known as Musiela parameterization \citep{musiela1993stochastic} and formally describes a future product $f_{t,m}$ as the price of the future product in $t$, with the corresponding delivery period starting in $t+m$.  Thus it holds for the time to maturity $m$ that $m = \min(\T)-t$.
If the delivery period is an interval $\T = [T_1, T_2]$, then we have $f_{t, m} = F_{t, \T }$ with $m= T_1-t$.
In the modeling section we consider the Musiela parametrization as well, as was also done for instance in \cite{barndorff2014modelling}, \cite{carmona2014survey} or \cite{benth2017space}.

To display the direct relationship of future products to expected prices, it helps to rearrange equation (\ref{eq_Risk_Premium}) to $\E[P_{\T}|I_t]$:
\begin{equation} \label{RP-eq}
\E[P_{\T}|I_t] = \text{RP}_{t,\T} + F_{t,\T} 
\end{equation}
It can be seen that there is a direct theoretical link between the expectation for electricity prices in the future, which can be e.g. generated by econometric modeling, and the price of the corresponding future product. Assuming that the risk premium is 0, we could easily obtain future electricity prices by taking a look at the future products. However, several authors have found various results concerning the risk premium, usually stating that there is a negative or positive risk premium present, usually determined by a complex set of variables see e.g. \cite{redl2013determinants} or \cite{aoude2016electricity}. Given historical information on day-ahead electricity prices and futures as well as other relevant information concerning the risk premia it is possible to construct and forecast the hourly price forward curve. This was done with real data for the German and Austrian electricity market for instance by \OLD{Paraschiv et al. (2015) and} \cite{caldana2017electricity}. Even though the authors had to forecast electricity spot prices as well, the focus of their study was to get realistic approximations for the hourly price forward curve. \NEW{\cite{paraschiv2015spot} utilized the estimated hourly price forward curves to simulate realistic hourly day-ahead spot price behavior for the German/Austrian market. They also conduct a forecasting study with two different time points with two different forecasting horizons each. They show that their combined regime-switching approach yielded better results than a combined ARIMA benchmark when the mean absolute percentage error (MAPE) is considered. Due to the nature of their approach and the fact that they have a similar forecasting horizon as we do, we will compare our models later on in detail.}  \NEW{However,} our model will differ in the sense that we focus on capturing the day-ahead price movements only by using the observable historic futures, for which we do not necessarily need the full hourly price forward curve. Nevertheless, the possible dependency of day-ahead electricity prices on future products is rather complex and needs a specific modeling approach.

\OLD{An approach which can capture the information of different future products might therefore tackle the issue of stacking errors of forecasted regressors. From a finance perspective we would simply utilize the market expectations for future prices to improve our forecasts. Assuming that we have a different information set than active traders, e.g. about actual outages or maintenances of power plants, this is a promising approach to improve our forecasts. Therefore, we structured our paper as follows. The next section will describe in detail how to merge both quite different market structures into one model. We will propose a model with an efficient estimation and regressor selection algorithm to gain high forecasting precision. In Section 3 we will execute a thorough forecasting study analyze our findings minutely. The last section will conclude by summarizing our results and pointing out the drawbacks of our study as well as suggestions for future research.}

\NEW{We will therefore setup a model which will use future products observable at a time point $t$ to forecast hourly day-ahead prices of up to four weeks with an econometric modeling. Our model does not need to explicitly create hourly price forward curves but instead will model the impact of future product prices directly towards the day-ahead electricity prices. Together with additional known regressors like the weekday or the seasons of the year we make sure that every of our external regressors is deterministic throughout the whole forecasting period. Such an approach, which can capture the information of different future products, might therefore tackle the issue of stacking errors of forecasted regressors. From a finance perspective this coincides with simply utilizing the market expectations for futures to improve our forecasts. Assuming that we have a different and especially worse information set than active traders, e.g. about actual outages or maintenances of power plants, this is a promising approach to improve forecasts. Therefore, we structured our paper as follows. The next section will describe in detail how to merge both quite different market structures into one model. We will propose a model with an efficient estimation and regressor selection algorithm to gain high forecasting precision. In Section \ref{sec_forecasting} we will execute a thorough forecasting study to analyze our findings minutely. The last section will conclude by summarizing our results and pointing out the drawbacks of our study as well as suggestions for future research.}

\section{Data and model setup} 
\begin{figure}[h!]
\resizebox{1.0\textwidth}{0.3\textheight}{\includegraphics[scale=1.0]{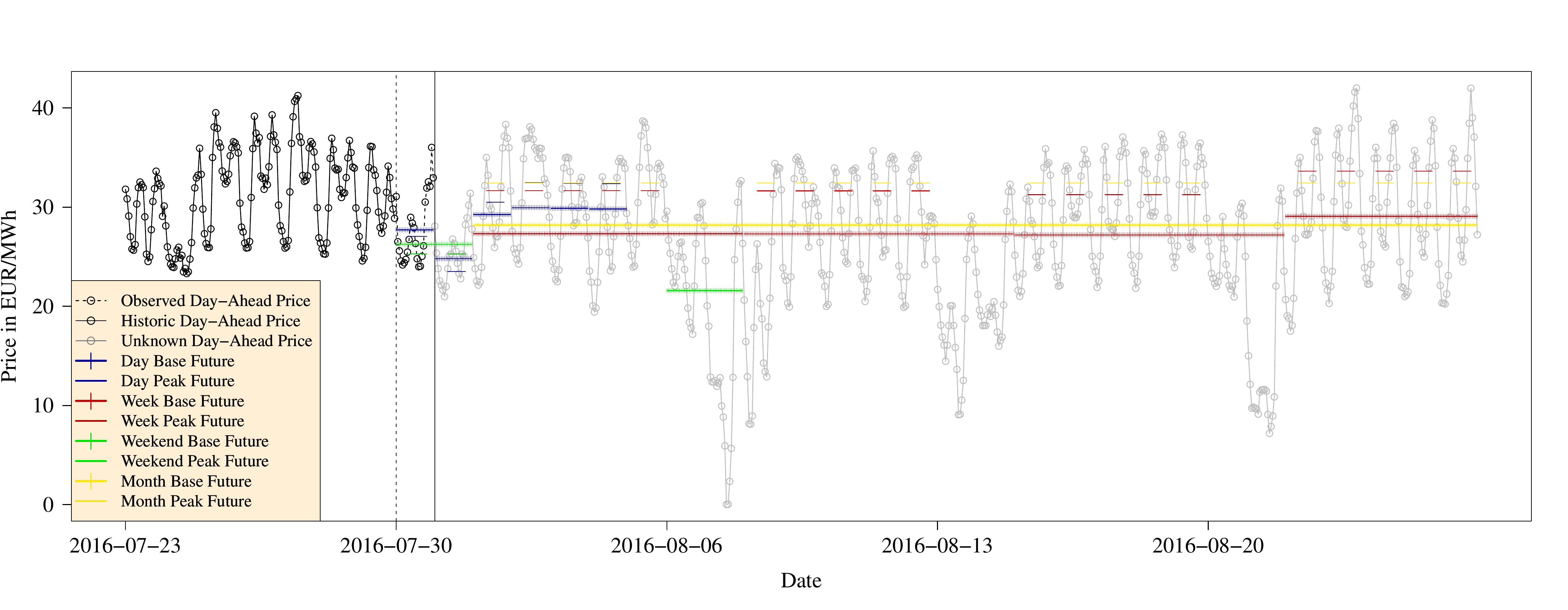}}
\caption{The day-ahead electricity price and the considered future products of our model. The different colored lines show Future products traded at the end of 29.07.2016 aligned to the time to which their underlying corresponds. Vertical lines depict the difference of one day for which the day-ahead price is known in the future.}
\label{fig_price_fut}
\end{figure}

Figure \ref{fig_price_fut} shows the complexity of the different future products and the day-ahead electricity price. We selected the future products, which we utilized in this paper and plotted their settlement price of the 29.07.2016 corresponding to the time frame they were traded for. For instance, as the 29.07.2016 was a Friday, the weekend base future with maturity one depicted as green line referred to the day-ahead electricity price of all hours of the weekend, e.g. the 30.07. and 31.07. The futures are based on end-of-day prices, meaning that because of the market structure of the spot products, the day-ahead electricity prices for the 30.07. were actually observable. This situation is depicted by the \NEW{two different vertical lines}\OLD{dashed vertical line which represents the change from 29.07. to the 30.07. and the vertical solid line which represents the point from which on the day-ahead price was not observable anymore}. Based on the observable day-ahead prices, the traders had to determine prices for different future periods by making predictions for the future electricity price, which is depicted as grey line. \OLD{This decision making process ends in the different product prices with different maturities which we depicted as various colored lines.} 

It can be easily seen that the traded future products of the 29.07. contained a great amount of information for the next several days, but due to week and month future also some information regarding the next four weeks. This is remarkable, as all this information is deterministic \NEW{and therefore a model which incorporates futures as regressors would not have the need to forecast these regressors as well.} \OLD{meaning that if it can be shown that these values actually influence the day-ahead price, we can use them to forecast the day-ahead electricity price directly and therefore avoid, to some extend, the uncertainty  forecasting. }

As mentioned in the previous section econometric models usually tend to have superior forecasting ability in the short-run with hourly precision, but struggle to make promising forecasts when forecasting horizons of weeks are considered. On the other hand future data provide deterministic long-term forecasts for horizons up to years but lack in accuracy when hourly or even to some extent even daily data is considered.

Hence, we are going to combine the precision of these econometric models with the power of the long-term price information of futures. We will setup an econometric model with a forecasting horizon of up to four weeks for the day-ahead electricity price of Germany and Austria while still preserving hourly precision.

Our model will therefore consist of three parts: the autoregressive time-series of day-ahead prices of the EPEX Spot for Germany and Austria, specifically selected future Phelix data for Germany and Austria provided by the EEX and day of the week dummy variables. Starting with the \OLD{01.02.2016} \NEW{01.05.2016} we will create a forecast for up to four weeks for every day of one year, which will result in \OLD{366} \NEW{365} forecasts each with a time horizon of up to 28 days. Every day where an estimation and forecasting is done will only use observations from the previous 365 days, meaning that we will utilize a rolling window for the estimation. 

We will use the EEX end-of-day data from the Phelix Base Day, Phelix Base Week, Phelix Base Weekend and Phelix Base Month products and their peak counterparts. We generally selected the different products and maturities to match our forecasting horizon. As second criterion we only incorporated products which were traded on business days for at least 75\% of the time. This was used as some products, for instance the weekend peak future from maturity 8 to 12 seem to have very little trades during our investigated time period. Third, for products with more than 10 different traded maturities, we decided to investigate only the most up-to-date maturities for which the delivery period of the product corresponded to time of the day-ahead product. For instance, the month future for the next month has maturities from 3 to 31, but only the maturity 3 contains the most recent information about the price for the next month. A more detailed discussion of the update scheme will be provided after the model is fully introduced.

It is important to notice that these datasets only update once a day, as we use end-of-day data. Also the delivery period of these products are defined by their specific type, e.g. the Phelix Base Week will deliver 1 MW/h for every hour of the specified week for the traded price via cash settlement.

Another issue occurs when trading times are considered. The future products are traded from 08:00 to 18:00 but as we use the end-of-day data our datapoints will always represent the price of 18:00. The day-ahead auction for the spot closes 12:00 and sets the prices for the following day. Given that situation the day-ahead prices can only be dependent on the future end-of-day prices of two days before the day for which they were traded. Unlike the day-ahead electricity the future products are only traded on business days, so there is no price available on weekends or public holidays. However, our model accounts for all these facts. In the following description of the model parts we will mention the specific adjustments to overcome these issues.

The full model is defined as follows:
\begin{align}
Y_{d,h} =&  \underbrace{ \sum_{j=1}^{24} \sum_{k=1}^{K_Y} \beta_{h,j,k} Y_{d-k,h}}_{\text{spot prices}} 
+ \underbrace{ \sum_{k=0}^{K_{\text{day}}} \sum_{m=2}^6 \beta_{f^{\text{day}},h,k,m} f_{d-k,m}^{\text{day}} }_{\text{base day futures}}  
+ \underbrace{ \sum_{k=0}^{K_{\text{week}}} \sum_{m \in M_{\text{w}}} \beta_{f^{\text{week}},h,k,m} f_{d-7k,m}^{\text{week}} }_{\text{base week futures}}  \nonumber \\
&+ \underbrace{  \sum_{k=0}^{K_{\text{wkend}}} \sum_{m \in M_{\text{wk}}} \beta_{f^{\text{wkend}},h,k,m} f_{d-7k,m}^{\text{wkend}}  }_{\text{base weekend futures}} 
+ \underbrace{  \beta_{f^{\text{month}},h} f_{d,1}^{\text{month}} \vphantom{\sum_{k=0}^{K_{\text{wkend}}}}  }_{\text{base month future}}  \nonumber \\
&+ \underbrace{ \sum_{k=0}^{K_{\text{day}}} \sum_{m=2}^6 \beta_{f^{\text{day},\text{p}},h,k,m} f_{d-k,m}^{\text{day},\text{p}}  }_{\text{peak day futures}} 
+ \underbrace{ \sum_{k=0}^{K_{\text{week}}} \sum_{m \in M_{\text{w}}} \beta_{f^{\text{week},\text{p}},h,k,m} f_{d-7k,m}^{\text{week},\text{p}}  }_{\text{peak week futures}}   \nonumber \\
&+\underbrace{ \sum_{k=0}^{K_{\text{wkend}}} \sum_{m \in M_{\text{wk},\text{p}}} \beta_{f^{\text{wkend},\text{p}},h,k,m} f_{d-7k,m}^{\text{wkend},\text{p}}  }_{\text{peak weekend futures}} 
+ \underbrace{ \beta_{f^{\text{month},\text{p}},h} f_{d,1}^{\text{month},\text{p}} \vphantom{\sum_{k=0}^{K_{\text{wkend}}}} }_{\text{peak month future}} 
+ \underbrace{ \sum_{k=1}^7\beta_{k,h} \text{DoW}_k }_{\text{weekdays}} +\underbrace{ \sum_{k=1}^4\beta_{k,h} \text{S}_k }_{\text{periodic B-splines}} 
 + \epsilon_{d,h} 
\label{main_model}
\end{align}

The model\OLD{, which} represents an AR24-X model\NEW{, where X stands for the external regressors, e.g. futures, weekday dummies and periodic B-splines. It contains the regressor coefficients $\beta$ as well as an error term $\epsilon_{d,h}$. The model}\OLD{ with coefficients $\beta$, and error term $\epsilon_{d,h}$,} treats the day-ahead electricity price $Y_{d,h}$ for every hour $h$ of the day $d$ as a separate time series. This is useful especially as future data is usually not available in hourly precision but can be e.g. for the day-base-future in daily precision. As the day-ahead prices for each hour now also exhibits daily precision the variations of both time series occur within the same time period. We will refer to model \eqref{main_model} as Future-Model. 

\begin{figure}[h!]
\resizebox{0.95\textwidth}{0.67\textheight}{\input{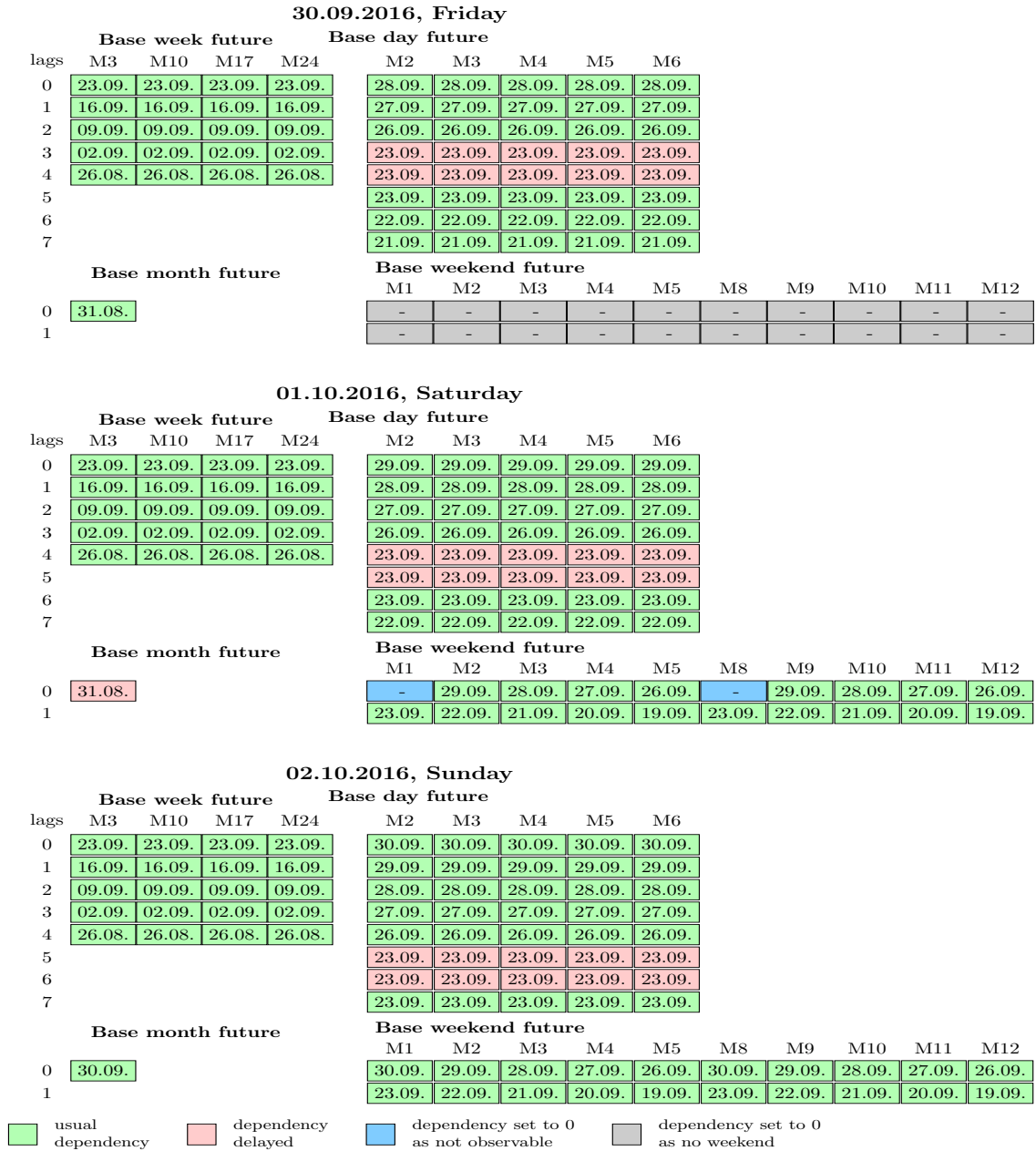}}
\caption{Illustration of the base future product dependency structure for the day-ahead electricity price of three distinct days, entries show the date at which the product was traded, M stands for the maturity of the product.
}
\label{fig_illustration_dep}
\end{figure}

Every hourly price of a day in the dataset is dependent on the previous $K_Y=7$ days of itself as well as on the 23 other hours of a day and their previous $K_Y=7$ lags to model the autocorrelation structure. A detailed example for the dependency structure is presented in Figure \ref{fig_illustration_dep}, which should be kept in mind when reading the following technical descriptions of the future product dependency.

The variables $f^{\text{day}}$ and $f^{\text{day},\text{p}}$ represent the the prices of the Phelix base and peak day futures respectively for maturity $m$ of \OLD{1} \NEW{2} up to 6 days, based on the end-of-day price of the day two days before the actual observed day-ahead price. The price of 30.09.2016 e.g. is explained by all future maturities as end-of-day data of the 28.09.2016 as can be obtained from Figure \ref{fig_illustration_dep}. Higher maturities starting from up to 7 days in the future were not included as during the sample period no or just little trading occurred for these maturities. Additionally, the $K_{\text{day}}=7$ earlier prices of these maturities are considered as well. This guarantees that at least for one week we can use deterministic values, as for instance the day-base future with maturity 4 traded on Monday has now influence on the day-ahead price of Friday by the variable $f_{d-2,4}^{\text{day}}$, e.g. the maturity 4 base day-future price with lag 2. Notice that in this example the day-future with maturity 4 traded on Monday represents the price an investor had to pay on Monday to buy 1 MW/h of electricity for every hour on Friday. If the exchange was closed due to a public holiday or weekend we simply replaced the not available value with the most recent value for that maturity, of usually one or two days before that day. 

The data is also dependent one the last observable value of $f^{\text{week}}$ and $f^{\text{week},\text{p}}$ of the base and peak week future respectively, as traded on the antecedent week for the actual week. The last observable week future price for the current week is usually the one on the Friday of the last week, but can be some days before that when public holidays occurred. So the day-ahead price for every hour of, for instance, the 22.10.2016, which is a Saturday, is dependent on the end-of-day price of the week-future of 14.10.2016, which is a Friday. The reason for the inclusion of this value is that traders may use this value as an orientation for the mean price of the upcoming week. It is also a day of the week to which the traded week future at the 14.10.2016 corresponded to. Moreover, we included all traded maturities on Friday for the next four weeks, which are represented by the index set $M_{\text{w}} = \{3,10,17,24\}$ as well as their historic values with up to $K_{\text{week}}=3$ lags. This guarantees that we have deterministic values for up to four weeks, as $f_{d-3\times7,4}^{\text{week}}$ represents the value for the current week as traded on a Friday four weeks ago. Higher lags and maturities were dropped as they would exceed our planned forecasting horizon.


Furthermore, we included the base and peak futures for the weekend $f^{\text{wkend}}$ and $f^{\text{wkend},\text{p}}$ respectively. This product is traded only with maturities up to 12 days, which means that it only provides deterministic values up to the weekend of the following week. For our model we used every tradeable maturity, from 1 to 5 days-ahead for the current weekend and from 8 to 12 days-ahead for the following weekend. These maturities are represented by the index set $M_{\text{wk}}$. As the maturities 8 to 12 for the weekend peak future were traded very rarely or even never, we excluded these five time series from our model and labeled their corresponding set of maturities $M_{\text{wk},\text{p}}$. For the used weekend series we rearranged the prices so that for every day of the business week, e.g. Monday to Friday, every series becomes 0. The weekend days received the observable prices of every maturity traded during the week. This means that for instance the day-ahead prices for every hour of Sunday are dependent on the weekend base future price for maturity 1, which is always traded on a Friday, but also for maturity 2, which is always traded on a Thursday and so on. Note that the price for Saturday cannot be dependent on the price for maturity 1, as we used end-of-day data for the future, which results in that price being not possible to observe for investors. Hence, this specific price is set to 0 as well.

The last future product we included is the month base and peak future $f^{\text{month}}$ and $f^{\text{month},\text{p}}$ respectively. The value of the month future was calculated as the last traded and observable month future price of the last month for the current month. Note that the last observable price is not necessarily the one at the last day of the month, as due to weekends and public holidays the last day of a month may not be a trading day. In this case we instead took the first observable price before that day. The month future price was kept fixed over the whole month and changes only after a new month occurs. Due to our forecasting horizon we decided to not include any historic values of this product.

Finally, we also introduced seven dummy variables $\text{DoW}_k$ which represent the day of the week ($\text{DoW}$) and are set to 1 if a specific weekday occurs and are 0 otherwise. \NEW{To account for public holidays we regarded each public holiday as a Sunday. In addition to the weekday effects we also added the typical seasonal structure of electricity prices by adding a periodic cubic B-spline $S$ for each of the four season of the year. This regressor is a smooth cubic function which peaks at the the middle of a season and smoothly diminishes into the next season until it reaches the value of 0 for all other dates until the same season is about to start again. For details on the construction of periodic B-splines we refer the reader to \cite{ziel2016wind}.}

As already mentioned the rather complex dependency structure of the day-ahead price towards the future products is illustrated for an interesting set of days from 30.09.2016 to 02.10.2016 in Figure \ref{fig_illustration_dep}.\footnote{For simplification, we present the dependency structure only towards base future products and leave out the peak products. However, the only difference in the dependency structure between those two is that the maturity 8 to 12 weekend products were omitted for peak products.} These days show an interesting behavior as the include not only the phase in from a business day to the weekend but also the transition from one month to another. The different cells in the figure represent the day a specific product was traded, which is also the day for which the end-of-day value of that product was used as a regressor for the day-ahead-price of the day which can be obtained from the respective caption. For instance, for the day ahead-prices of 30.09.2016, which was a Friday, we can see that the base week future with maturitiy 3 (M3) traded at the 23.09. was used as a regressor. Moreover, it is shown that e.g. the base weekend future was set to 0 for that day, as Friday is not a day of the weekend. It is also identifiable, that for the base day future the values for 24.09. and 25.09. were not observable, as this was a weekend where the exchange was closed. Therefore the most recent values were used, which was in this case the value of the 23.09. which is a Friday. By comparing the different dates in the cells between the three presented days, one can retrace the distinct updating schemes as explained in the paragraphs before.

Overall, our model consists of \NEW{323} \OLD{335} possible parameters. 168 of them emerge from the 24 hours and their 7 lags. The day of the week dummies \NEW{together with periodic B-splines} account for 11 parameters. The remaining \OLD{160}\NEW{144} considered parameters coming from the future products are summarized in Table \ref{overview_fut}.

\begin{table}[h!] 
\resizebox{1.0\textwidth}{!}{
\begin{tabular}{ c| c |c| c| c| c |c| c| c}
& Day Base  & Day Peak  & Week Base  & Week Peak& Weekend Base  & Weekend Peak  & Month Base  & Month Peak   \\
& $f^{\text{day}}$ & $f^{\text{day},\text{p}}$ & $f^{\text{week}}$ & $f^{\text{week},\text{p}}$&$f^{\text{wkend}}$ & $f^{\text{wkend},\text{p}}$& $f^{\text{month}}$& $f^{\text{month},\text{p}}$ \\ \hline
Current Values & 1 & 1 & 1 & 1 & 1 & 1 & 1 &1 \\
Historic Values & 7 & 7 & 3 & 3& 1 & 1 & 0 & 0  \\
Number of Maturities & 5 & 5 & 4 & 4& 10 & 5 & 1 & 1 \\
Overall \# of parameters & (1+7)$\times$5=40 & (1+7)$\times$5=40 & (1+3)$\times$4=16 & (1+3)$\times$4=16 & (1+1)$\times$10=20 & (1+1)$\times$5=10 & (1+0)$\times$1=1 & (1+0)$\times$1=1
\end{tabular}
}
\caption{Overview about used future-products with their respective maturities and lags}
\label{overview_fut}
\end{table}

As our observation window has only 365 points in time for every hour and \NEW{323} \OLD{335} possible parameters, we need an efficient and sparse method to estimate and possibly eliminate some of the regressors driven by an algorithm. Hence, we decided to use the lasso estimator of \cite{tibshirani1996regression} in combination with a coordinate-descent estimation approach as included in the R-package \textit{glmnet} by \cite{friedman2016glmnet}. \NEW{For more insights into the lasso procedure and its properties, see e.g. \cite{hastie2015statistical}.}

As the lasso estimator is a penalized least square estimator, we need to rewrite equation \eqref{main_model}. Furthermore, in order to successfully execute the lasso algorithm the variables need to be standardized, so that they have a variance of 1. The final lasso-suited ordinary least squares representation of \eqref{main_model} is therefore:

\begin{align}
\wtilde{Y}_{d,h} &=  \wtilde{\X}_{d,h} \wtilde{\bsbeta}_{h} + \wtilde{\eps}_{d,h}, 
\label{eq_model_ols_scale}
\end{align}
where the $\sim$-symbol represents the scaled versions of the respective data vector. With representation \eqref{eq_model_ols_scale} we can estimate the scaled parameter vectors $\wtilde{\bsbeta}_{h}$ given all observable days by using the lasso estimator $\what{\wtilde{\bsbeta}}_{h}$:
   \begin{align}
   \what{\wtilde{\bsbeta}}_{h} &= \argmin_{\bsbeta \in \R^{p_{h}}} 
\sum_{d=1}^n (\wtilde{Y}_{d,h} -  \wtilde{\X}_{d,h} \bsbeta)^2
+ \lambda_{h} \sum_{j=1}^{p_{h}} |\bsbeta_{j} | 
\label{eq_lasso}
   \end{align}
where $\lambda_{h}\geq 0$ is a penalty parameter, $n$ is the amount of observations used and $p_h$ the number of possible parameters. 

As estimation algorithm we use the
coordinate descent approach of \cite{friedman2007pathwise} especially as it provides a fast estimation technique. Given a possible parameter set $\Lambda_{h}$ estimated by the coordinate descent approach. We select the best fitting tuning parameter $\lambda_{h}$ on an exponential grid ($2^{\G}$ with $\G$ equidistant) by evaluating the Bayesian information criterion (BIC) which is regarded as conservative and therefore sparse model selection criterion. Using the BIC criterion will therefore induce a high probability that a large amount of the possible regressors will be removed by setting their equivalent $\beta$ to 0, so that the final models for every hour are very unlikely to exhibit all parameters. To keep track of the included regressors we will show an overview about how often they were actually used in the following section.

\NEW{Using this complex approach of modeling the data allows us to incorporate information in the day-ahead prices, which are not represented in the models which use the hourly price forward curves. Given the future price information of a specific day it is possible to create the hourly price forward curve (HPFC) as an risk-neutral approximation for the real day-ahead prices of the upcoming days. In order to get the real prices under the assumption that investors are risk-averse it is possible to model the differences of the hourly price forward curve to the real day-ahead prices as a risk premium. This was done for instance by \cite{paraschiv2015spot}. They use the future base prices of a specific day to create the HPFC in order to get information of the day-ahead prices of the future. To create realistic day-ahead prices under risk-averse evaluation they additionally model the difference between the two as risk premium using a regime switching time series approach. In this sense they are not only able to track the movements of the risk premium, but also to guarantee price spikes when they simulate prices with their model. As the time horizon of the future base products is relatively high, they can even carry out a forecasting analysis for two different starting time points in 2012 for up to one month.

Our approach is comparable to the model of \cite{paraschiv2015spot} as we use a similar set of input parameters to estimate day-ahead prices of the future. However, there is a substantial difference in our approaches, as in addition to the futures of one day, we also utilize the futures of days before today as lagged data. This has the advantage that we can use the historic expectations of investors who, for instance, have bought three days ago a day base future with maturity in four days. Such investors are obliged to pay or receive the settlement price determined by the future product. If these traders are active on the day-ahead exchange as well, they are likely to remember the price they had agreed on several days ago when they contracted the future product and may therefore influence the day-ahead price accordingly. 

Our approach would therefore roughly correspond to a HPFC-model which uses the HPFC of today and the past HPFCs as well as direct input to day-ahead prices. As we do not model HPFC directly, our model does not has to be in line with theory about risk averse investors and their required risk premia. Nevertheless, if the assumptions of the theory leading to equation \eqref{RP-eq} are correct, HPFC-approaches and our approach model the same variable,}  \NEW{i.e. $\E[P_{\T}|I_t]$. If we further assume that the risk premium is exactly 0, then our model would also model the HPFC in a comparable fashion as \cite{caldana2017electricity}, who also utilize a function of day-ahead prices with e.g. seasonal and weekly patterns to create HPFC.

However, even though our regressors can be motivated by the theory of HPFC, we do not necessarily require the underlying assumptions of that theory to hold. Our model is only associated with the assumptions of plain time-series analysis. The reason for that is that our model has the sole purpose of forecasting day-ahead prices, which is also not necessarily the same as explaining the underlying data, see for instance \cite{shmueli2010explain} for a thorough analysis of that common misconception. This insight leads to another important difference in our modeling approaches, as we utilize the advantages of machine learning for parameter selection, as done by the estimation algorithm of \cite{friedman2007pathwise} for the lasso-selection problem we use. Therefore we can simply add more regressors, e.g. deterministic information on the weekday or previous lags of the price. 

However, our approach has the drawback compared to the approach of \cite{paraschiv2015spot}, that due to the lack of a financial theory for our model we cannot determine the risk premium which risk averse investors require.}

\section{Forecasting, results and simulation} \label{sec_forecasting}

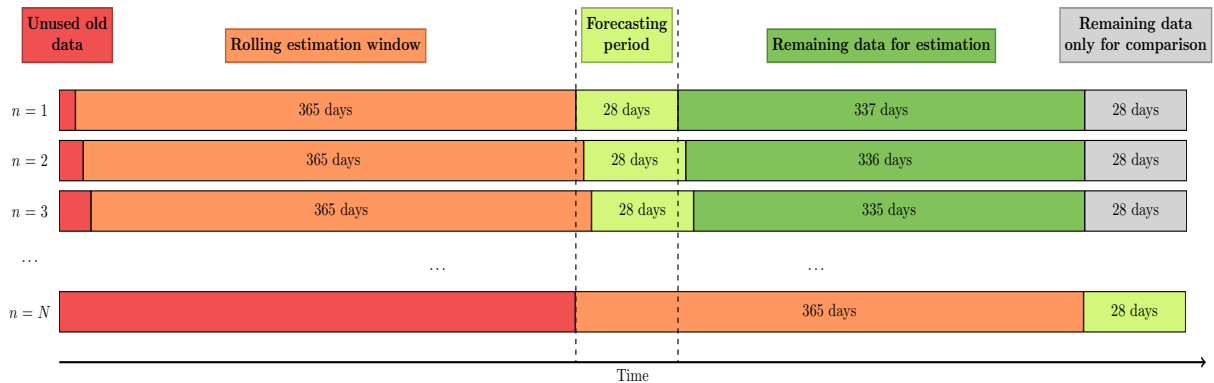
\begin{figure}[h!]
\resizebox{1.0\textwidth}{0.2\textheight}{\begin{tikzpicture}[node distance=0pt]
\TimeLine{%
    \the\numexpr\mydaylength/red1/{},%
    \the\numexpr\mydaylength*32/yellow2/{365 days},%
    \the\numexpr\mydaylength*6.5/green1/{28 days},%
    \the\numexpr\mydaylength*26/green2/{337 days},%
    \the\numexpr\mydaylength*6.5/grey1/{28 days}%
  }
\AddText{red1}{Unused old \\ data}{1}{M}
\AddText{yellow2}{Rolling estimation window}{2}{M}
\AddText{green1}{Forecasting \\ period}{3}{M}
\AddText{green2}{Remaining data for estimation}{4}{M}
\AddText{grey1}{Remaining data \\ only for comparison}{5}{M}
\coordinate (dash) at (part2.east);
\coordinate (dash2) at (part3.east);
\coordinate (p1) at (part0);

\TimeLinetwo{%
    \the\numexpr\mydaylength*1.5/red1/{},%
    \the\numexpr\mydaylength*32/yellow2/{365 days},%
    \the\numexpr\mydaylength*6.5/green1/{28 days},%
    \the\numexpr\mydaylength*25.5/green2/{336 days},%
    \the\numexpr\mydaylength*6.5/grey1/{28 days}%
  }
  \coordinate (p2) at (part0);
  
  \TimeLinethree{%
    \the\numexpr\mydaylength*2/red1/{},%
    \the\numexpr\mydaylength*32/yellow2/{365 days},%
    \the\numexpr\mydaylength*6.5/green1/{28 days},%
    \the\numexpr\mydaylength*25/green2/{335 days},%
    \the\numexpr\mydaylength*6.5/grey1/{28 days}%
  }
  \coordinate (p3) at (part0);
      \TimeLineinvis{%
    \the\numexpr\mydaylength*71.5/red1/{}%
  }
  \coordinate (p4) at (part0);
  
    \TimeLinefour{%
    \the\numexpr\mydaylength*33/red1/{},%
    \the\numexpr\mydaylength*32.5/yellow2/{365 days},%
    \the\numexpr\mydaylength*6.5/green1/{28 days}%
  }
  \coordinate (p5) at (part0);
  
  \draw[dashed,thick] ([yshift=2cm]dash) -- ([yshift=-5cm]dash);
  \draw[dashed,thick] ([yshift=2cm]dash2) -- ([yshift=-5cm]dash2);
  \draw[->, very thick] ([yshift=-1cm]part0.west) -- ([xshift=0.5cm,yshift=-1cm]part3.east)node [pos=0.5, below] {Time}node [pos=0.33, above=1.7cm] {$\ldots$}node [pos=0.66, above=1.7cm] {$\ldots$};
  \node[left=of p1,align=center,text width=1.2cm] {$n=1$};
  \node[left=of p2,align=center,text width=1.2cm] {$n=2$};
  \node[left=of p3,align=center,text width=1.2cm] {$n=3$};
  \node[left=of p4,align=center,text width=1.2cm] {$\ldots$};
  \node[left=of p5,align=center,text width=1.2cm] {$n=N$};

\end{tikzpicture}}
\caption{Depiction of the used rolling window scheme for our forecasting analysis.}
\label{fig_rolling}
\end{figure}

\NEW{
We design a rolling window out-of-sample forecasting study for forecasting evaluation.
Note that in contrast to an expanding window study only a rolling window study allows for proper forecasting evaluation in the context of the Diebold-Mariano test that we utilize here, for more details see \cite{diebold2015comparing}.
In our study we consider an in-sample length of $365$ days.
}
\OLD{We design a rolling window out-of-sample forecasting study so that we have always the same fixed in-sample length of $366$ days.}
Moreover, we use $N=365$ rollowing windows. We denote $D$ as the last in-sample day of the first rolling window.
Hence, we define the estimated error $\what{\epsilon}_{c,h,n}$ of model \eqref{main_model} as $c$ days and $h$ hours ahead forecasting error of the $n$-th rolling window.
Note that  $\what{\epsilon}_{c,h,n}$  is an estimated error of $\epsilon_{D-1+ n+c,h}$, but $\epsilon_{D-1 +n+c,h}$ is estimated by multiple models as the forecasting windows overlap within the rolling window study. Figure \ref{fig_rolling} illustrates this design of the forecasting study. Reddish colors symbolize data which was observed until the start of the forecasting period. Greenish colors stand for data which originates from the future. The greenish area consists of the forecasting period and the remaining data for estimation and covers one year, e.g. $28+337=365$ days. For $n$ from 1 to 3 it can be seen that the rolling estimation window shifts for one day for each $n\in N$. This is done until $n=N$, e.g. the whole greenish area of $n=1$ is covered. The dashed lines show the range of the forecasting period of $n=1$ and helps to determine the structure of the rolling process. It can be obtained that the day of the last forecasting period of $n=1$ corresponds to the day of the next to last day of the forecasting period of $n=2$ and so on.

After estimating the coefficients of model \eqref{main_model} we can simply construct a forecast by iteratively estimating 
$Y_{D-1+n+c,h}$, where $c \in C_{\text{week}}=\{1,2,3,\ldots, c_{\max}\}$ 
represents the index set of days given the starting weekday which has to be forecasted \NEW{ and $c_{\max}$ of $C_{\text{week}}$ is 28}. \NEW{However, some of the future products are only observable for some of the days of the prediction horizon. E.g. the day base future with maturity 2 is only observable for the day-ahead price in two days. As forecasting these future products would lead to a situation where we would still had to face the problem as described earlier, we estimate the coefficients in model \eqref{main_model} repeatedly for each day of our 28 days long forecasting horizon, but exclude all regressors which were not observable for that time. This has the advantage that we will only use true observable and therefore deterministic regressors, which makes the model fair and applicable for practical uses. The disadvantage is that we have to carry out 28 out model estimations per shift in the rolling estimation window, which coincides with demanding CPU-times.} \OLD{This is always every day up to four weeks, so 
the maximal element $c_{\max}$ of $C_{\text{week}}$ is 28.  The forecasting algorithm automatically checks whether a non-autoregressive component was observable on the starting day and then uses them either deterministically when it was observable or forecasts the regressor itself when it was not. This forecasting is done by a very simple AIC-based iterative AR(p). By doing this we ensure that the model remains practical and especially fair, as we use only data which in reality was observable at that given point in time.}

For model comparison we decided to use two different and commonly used measures, the MAE (Mean Absolute Error) and the MMAE (Mean Mean Absolute Error). These two are defined as follows:

\begin{align}
\MAE_{c,h}& =\frac{1}{N}\sum_{n=1}^{N} |\what{\epsilon}_{c,h,n}| \\
\MAE_{k} &=  \MAE_{\lfloor (k-1)/24 +1\rfloor, k \,\text{mod}\, 24 + 1} \\
\MMAE_{K}&=\frac{1}{K}\sum_{k=1}^{K}\MAE_{k} 
\end{align}
with $\lfloor \cdot \rfloor$ as smallest integer and $\text{mod}$ as modulo operator.

Given the $N=365$ forecasts of our rolling estimation window, which included forecasts for every day of the months from \NEW{May 2016 to April 2017} \OLD{February 2016 to January 2017}, we can compute the $\MAE_{c,h}$ which represents the Mean Absolute Error for every day $c$ and every corresponding hour of the day $h$. As we forecasted four weeks the $\MAE_{c,h}$ can be calculated for up to $24\times7\times4=672$ hours. Given a $\MAE_{c,h}$ we can easily transform it to the $\MAE_{k}$ notation, which is convenient to calculate the $\MMAE_K$.
The $\MMAE_{K}$ is a less volatile measure than the $\MAE_{k}$ as it represents the error the model has made up to a certain forecasting horizon in opposition to the $\MAE_{k}$ which provides the error for a specific forecasting horizon.

For model comparison we introduce three extremely competitive benchmarks, \OLD{the AR(p) with AIC-selection, which captures several types of autoregressive models 
when $p$ is sufficiently chosen, see e.g.
Ziel (2016) for more details on this topic, } 
\NEW{two VAR-HoW(p) with AIC-selection based on the paper of \cite{ziel2015efficient} } 
 \OLD{the AR-HoW(p) with AIC-selection based on the paper of} 
and the AR24(p) with lasso and BIC-selection as it was selected as most competitive model in \cite{ziel2016forecasting}. They are defined as follows:
\begin{eqnarray}
\NEW{ \text{VAR-HoW(p)'s}:} &\NEW{ \bsY_t} &= \NEW{ \overline{\bsY_{\text{HoW},t}} + \sum_{k=1}^{p} \bsB_k  (\bsY_{t-k} - \overline{\bsY_{\text{HoW},t}}) + \epsilon_t } \\ 
\text{AR24(p)}: & Y_{d,h} &= \sum_{j=1}^{24} \sum_{k=1}^{K_Y} \beta_{h,j,k} Y_{d-k,h} +\sum_{k=1}^7\beta_{k,h} \text{DoW}_k + \epsilon_{d,h} 
\end{eqnarray}
\NEW{For the} \OLD{The AR(p)} \NEW{VAR-HoW(p) models} \OLD{with AIC-selection the day-ahead electricity price $Y_t$ was treated as one univariate time-series, i.e. $Y_{24(d-1) + h} = Y_{d,h}$. }
\NEW{the $\bsY_t$ was treated as one hourly time-series, i.e. $\bsY_{24(d-1) + h} = \bsY_{d,h}$.}
\NEW{ The first one of the VAR-HoW(p) models represents a univariate model with $\bsY_t = Y_t$, which we refer to as AR-HoW(p) from now on.
The second one is a 3-dimensional model for the electricity price, electricity load and renewable energy production from wind and solar energy,  so $\bsY_t = (Y_t, L_t, R_t)$ with $L_t$ as load and $R_t$ as renewable energy production. This model will be referred to as VAR-X(p). Note that the univariate and bivariate models were competitive benchmarks 
in \cite{ziel2015efficient} for the same forecasting horizon.} 
 \NEW{The univariate AR-HoW(p) model in a slightly modified version turned out to be a strong competitor in the paper of \cite{ziel2016day}. }



\NEW{The VAR-HoW(p) extends the standard VAR(p) framework by a weekly seasonal component, so that the resulting process is non-stationary. The seasonal regressor HoW which stands for Hour of the Week represents a deterministic dummy variable which becomes 1 when a specific hour of the week is present. }
\OLD{The AR-HoW(p) extends the AR(p) by the regressor HoW which stands for Hour of the Week and represents a deterministic dummy variable which becomes 1 when a specific hour of the week is present. }
As a week has $7\times24=168$ hours in total this results \NEW{in 168 regressors added for each dimension of the VAR-HoW(p).}\OLD{in 168 regressors added in comparison to the AR(p).} The estimation of this model was done in a two-step-approach by first estimating the model
\NEW{ $\bsY_t = \sum_{i=1}^{168} \bsgamma_i  \text{HoW}^i_{t}\bsone + \bseps_t$ 
via Ordinary-Least-Squares and then an AIC-selected VAR($p$) by solving the multivariate Yule-Walker-Equations on $\bseps_t$
 where $p$ is selected out of $1,\ldots,p_{\text{max}}=168$. 
Note that $\overline{\bsY_{\text{HoW},t}}=\sum_{i=1}^{168} \bsgamma_i  \text{HoW}^i_{t}\bsone$. 
}
\OLD{ $Y_t = \sum_{i=1}^{168} \gamma_i  \text{HoW}^i_{t} + \varepsilon_t$ via Ordinary-Least-Squares and then applying an AIC-selected AR(p) as described before on $\varepsilon_t$. Note that $\overline{Y_{\text{HoW},t}}=\sum_{i=1}^{168} \gamma_i  \text{HoW}^i_{t}$. }

Our last benchmark is the AR24(p). It corresponds to our main model \eqref{main_model} but considers no future data. It is also estimated in the same way as our main model. With that benchmark we want to isolate the impact of the futures data towards our model.

The MMAE is a very good measure to break down different models to one number. This can be done by choosing the highest possible $K$, e.g. the mean over all forecasting errors up to $K=672$ hours in the future. After estimating and forecasting the day-ahead electricity price \OLD{with the same setting} for each model, we start with the presentation of the MMAE for a forecasting horizon of $K=672$ hours. It is displayed in Table \ref{MMAE_final}.
\begin{table}[h!]
\centering
\begin{tabular}{c|c|c|c|c}
&Future-Model&\NEW{AR24(p)}&AR-HoW(p)&\NEW{VAR-X(p)} \\ \hline
$\MMAE_{\NEW{672}}$& 7.19 & \NEW{8.56} & \NEW{8.26} & \NEW{8.49} 
\end{tabular}
\caption{MMAE for a forecast of four weeks for the final model and every benchmark model}
\label{MMAE_final}
\end{table}
In this first overview we can see that the proposed future-based model is superior considering the whole forecasting period. The number 7.19 means that a market participant who used this model faced an absolute error of 7.19 EUR/MWh on average, when the participant forecasted 672 hours or four weeks at once.

\begin{figure}[h!]
\resizebox{1.0\textwidth}{0.3\textheight}{\includegraphics[scale=1.0]{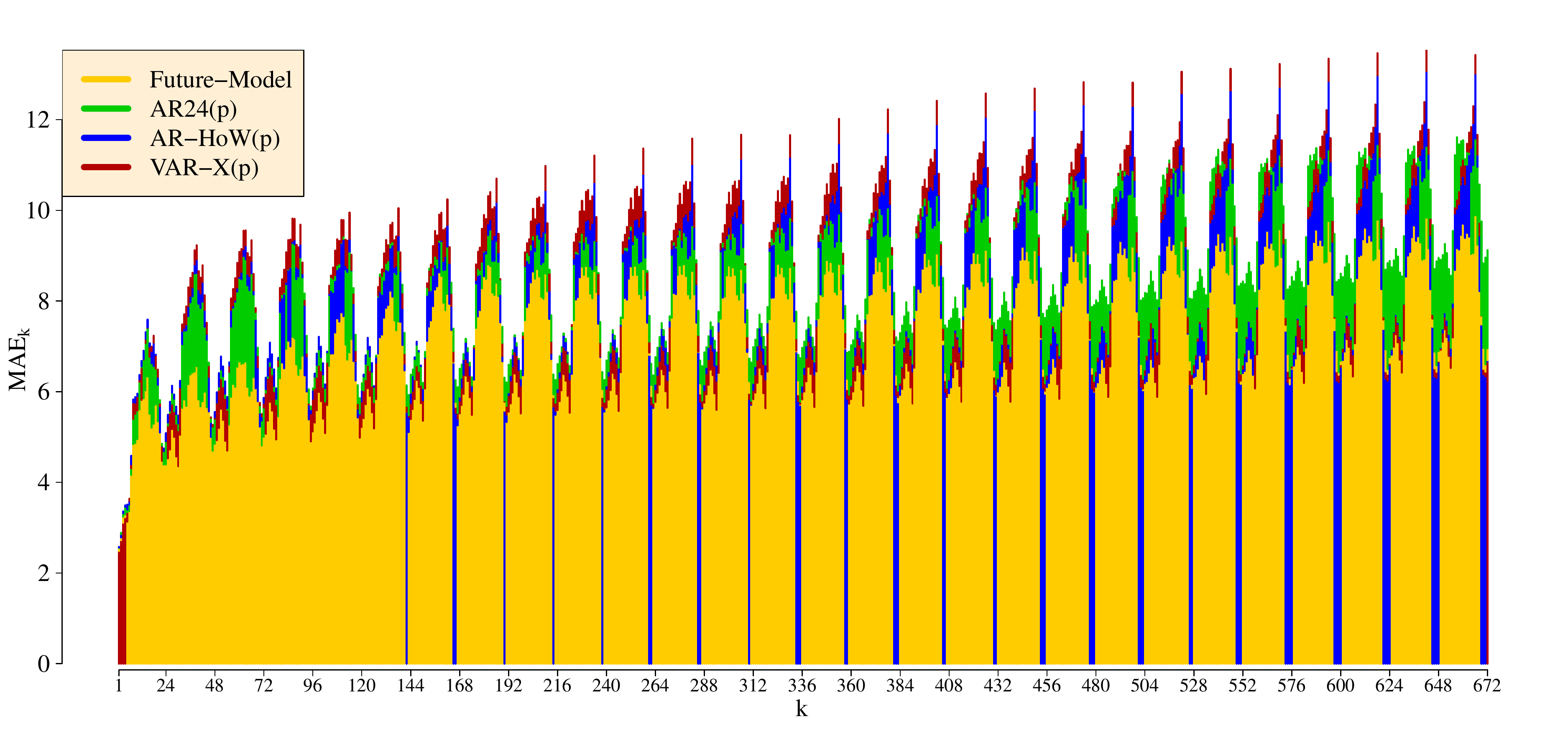}}
\caption{Stacked bar chart of the $\MAE_k$ for every model and every of the 672 forecasted hours. Bars for each hour are drawn so that the first bar represents the $\MAE_k$ of the best model and the bar stacked on that the difference to the anteceding model and so on. Colors thereby from which model the $\MAE_k$ was drawn.}
\label{fig_MAE_h}
\end{figure}

Much more detailed and difficult to illustrate is the $\MAE_k$ as we have to compare four different models over 672 points in time, resulting in $672\times4=2688$ different values. Hence, we decided to use a stacked bar chart over all 672 hours which shows the $\MAE_k$ for each model. This is shown in Figure \ref{fig_MAE_h}.
Every model received its individual color, as shown in the legend in the top-left corner of the figure. The stacked bars are always ordered from the best to the worst model, so that the viewer can easily spot the best model over the time. The difference of the next best model to the actual model is then stacked on the existing bar and given its individual color. This is done for every bar and every benchmark model so it is also possible to spot the worst model very quickly. Hence, it can be seen that the Future-Model which has the color yellow seems to outperform every other model over the majority of time. \NEW{Almost} each of the benchmark models is however for some hours in the future the best model. We can also see that the \NEW{VAR-X(p) with color red} \OLD{AR(p) with color blue} is very often the worst model for peak hours. \NEW{This is especially fascinating given the fact that for the first fours hours the VAR-X(p) is in fact the best model. The diminishing forecasting power of the model is very likely due to the fact that for this type of model the regressors, e.g. load and especially renewables have to be forecasted as well. But as these forecasts are barely accurate for renewables for such a long time frame they add uncertainty to the overall precision of the price model.} Interestingly, it can be obtained that every model independent of being univariate or multivariate, in the sense that it models every hour separately, tends to have extremely volatile $\MAE_k$ behaviour. The night hours seem to exhibit much lower $\MAE_k$ than the daytime hours. A possible reason for that could be the influence of renewables, as these hours are prone for unexpected variability of sunlight and the resulting solar power. As expected, the $\MAE_k$ also seems to increase over time, but reaches a convergence point after approximately \NEW{one week} \OLD{3 weeks}, especially when the Future-Model is considered. The best model during the first 24 hours of the forecasting horizon exhibits extremely low $\MAE_k$ values, always below a $\MAE_k$ of \NEW{7} \OLD{6} EUR/MWh. \OLD{Also the following days the Future-Model seems to be strictly better.} 

However, as there are still some hours where other model outshine the Future-Model we want to further investigate the differences between these models and determine which is the overall best model by using statistical theory. Therefore we employed a multivariate Diebold-Mariano-Test (DM-Test) for every day of the first four forecasting weeks. The DM-Test is a quite commonly used test in electricity price forecasting, it was recently used for instance in \cite{bordignon2013combining} or \cite{nan2014forecasting}.

Our DM-test is based on the $\|\cdot\|_1$-norm of the estimated daily residuals $\what{\boldsymbol{\epsilon}}_{c,n} = ( \what{\epsilon}_{c,1,n}, \ldots, \what{\epsilon}_{c,24,n}')$ 
of two forecasts, say $A$ and $B$.
It utilizes the loss functions $L_{c,n}^A = \| \what{\boldsymbol{\epsilon}}^A_{c,n} \|_1$ and $L_{c,n}^B = \| \what{\boldsymbol{\epsilon}}^B_{c,n} \|_1$ 
to compute the loss differences
$$ \Delta_{c,n}^{A,B} = L_{c,n}^A - L_{c,n}^B .$$
The key idea is now to check if $\Delta_{c,n}^{A,B}$ is significantly different from zero or not. If 
$\Delta_{c,n}^{A,B}$ is significantly smaller than zero with respect to a certain significance level then the forecast $A$ is significantly better than forecast $B$ for the forecasting day $c$.
The DM-statistic for forecasting day $c$ is defined as
$$\text{DM}_{c} = \frac{ \overline{\Delta}_{c}^{A,B} }{ \sigma(\overline{\Delta}_{c}^{A,B}) }$$
where $\overline{\Delta}_{c}^{A,B} = \frac{1}{N} \sum_{n=1}^N \Delta_{c,n}^{A,B}$ and 
$\sigma(\overline{\Delta}_{c}^{A,B})$ with its standard deviation. The latter one we estimate by the sample standard deviation of the corresponding process.
Note that under the null hypothesis of the test ($\text{H}_0: \Delta_{c,n}^{A,B} = 0$) the DM-statistic $\text{DM}_{c}$ is asymptotic normal.

\begin{figure}[h!]
\centering
\resizebox{0.8\textwidth}{0.2\textheight}{\includegraphics[scale=1.0]{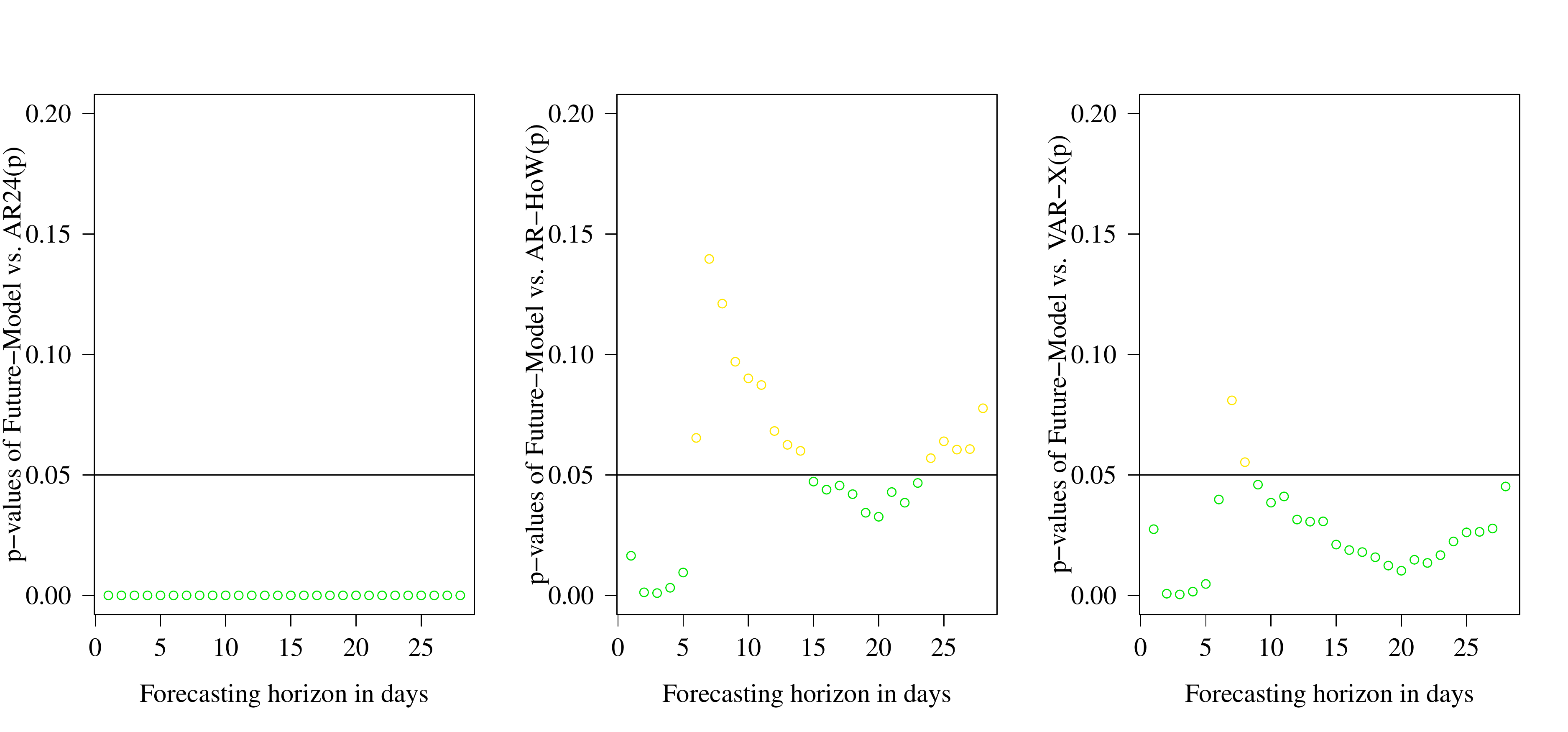}}
\caption{p-Values of the DM-Test for every day of the first 28 days of forecasting. Blue dots show days for which the Future-Model was significantly better than the competing model.}
\label{fig_DM}
\end{figure}

We applied the DM-Test for all competitors of our model, the \NEW{AR24(p), the AR-HoW(p) and the VAR-X(p)} \OLD{AR(p), the AR24(p) and the AR-HoW(p)}. The results are illustrated in figure \ref{fig_DM}. 
The figure depicts the p-values of the DM-Test for the Future-Model against all tested benchmark models. For every day of the first four weeks, e.g. 28 days, a multivariate DM-test was executed. The horizontal line in every chart represents \NEW{an arbitrarily} \OLD{a freely} chosen five percent error probability for our test. If one of the dots lays above the straight line, we can conclude that based on our level of confidence we cannot say that our model performed \NEW{significantly} better for that specific time horizon. \NEW{The figure indicates that for all investigated days the AR24(p) benchmark could not perform significantly better than our model. As the AR24(p) model is the same model as ours but without the futures, we can conclude that adding future products provided an increasing in forecasting performance. Even though overall the Future-Model performed better than the AR-HoW(p) and VAR-X(p) benchmark models, we cannot statistically proof that for every day of the forecasting. For the VAR-X(p) benchmark our model is significantly better for every day except for day seven and eight. Comparing our results with the AR-HoW(p) our models performs significantly better during 50 percent of all days. It can be obtained that especially during the first days our model exhibits great model performance, while the following days the performance seems to not become significant anymore. However, after day 14 the performance seems to stabilize again and the model becomes significantly better. Please note that our model not being significantly better does not mean that the benchmark models performed better. Also all these evaluations are heavily dependent on the chosen significance level. For a ten percent significance level for instance our model would be significantly better for almost every hour and model. But as these level choices, especially given a multiple testing problem, are overall debatable we will base our evaluation only for one significance level and provide the p-values in Figure \ref{fig_DM} at the ordinate.} \OLD{However, as the figure shows, there was no day, where our model did not significantly each of the competitors. Even though there were some hours as seen in Figure 4 were other models outperformed our model, it was not enough to detract the superiority of our model approach for the investigated time period.} Nevertheless, comparing simply the performance of models does not provide detailed insight to the reasons of why the Future-Model lead to superior forecasting performance. Hence, we would like to analyse the final model structure of our Future-Model throughout the whole forecasting period.

\begin{table}[h!]
\resizebox{0.95\textwidth}{0.5\textheight}{
\setlength\arrayrulewidth{1.5pt}
\begin{tabular}{cccccccccccccccccccccccccc}
  \hline
 && \textbf{1} & \textbf{2} & \textbf{3} & \textbf{4} & \textbf{5} & \textbf{6} & \textbf{7} & \textbf{8} & \textbf{9} & \textbf{10} & \textbf{11} & \textbf{12} & \textbf{13} & \textbf{14} & \textbf{15} & \textbf{16} & \textbf{17} & \textbf{18} & \textbf{19} & \textbf{20} & \textbf{21} & \textbf{22} & \textbf{23} & \textbf{24} \\ 
\multirow{48}{*}{\begin{sideways}\textbf{Day-Ahead Price}\end{sideways}}&H1 lag 1 & \cellcolor[rgb]{1,1,0.79} 0.00 & \cellcolor[rgb]{1,1,0.79} 0.00 & \cellcolor[rgb]{1,1,0.79} 0.03 & \cellcolor[rgb]{1,1,0.778} 1.06 & \cellcolor[rgb]{1,1,0.736} 4.67 & \cellcolor[rgb]{1,1,0.789} 0.10 & \cellcolor[rgb]{1,1,0.79} 0.00 & \cellcolor[rgb]{1,1,0.753} 3.15 & \cellcolor[rgb]{1,1,0.781} 0.74 & \cellcolor[rgb]{1,1,0.789} 0.07 & \cellcolor[rgb]{1,1,0.79} 0.00 & \cellcolor[rgb]{1,1,0.789} 0.08 & \cellcolor[rgb]{1,1,0.79} 0.00 & \cellcolor[rgb]{1,1,0.79} 0.03 & \cellcolor[rgb]{1,1,0.788} 0.17 & \cellcolor[rgb]{1,1,0.79} 0.00 & \cellcolor[rgb]{1,1,0.79} 0.00 & \cellcolor[rgb]{1,1,0.79} 0.00 & \cellcolor[rgb]{1,1,0.79} 0.00 & \cellcolor[rgb]{1,1,0.79} 0.00 & \cellcolor[rgb]{1,1,0.787} 0.25 & \cellcolor[rgb]{1,1,0.789} 0.08 & \cellcolor[rgb]{1,1,0.79} 0.00 & \cellcolor[rgb]{1,1,0.789} 0.06 \\
\hline
\end{tabular}

}
\caption{Illustration of the percentage of the actually used regressors for each of the 24 hour time series of the day-ahead price. The regressors for the day-ahead price from lag 2 to 6 were dropped from the illustration.}
\label{tab_BBTRUE}
\end{table}

\begin{table}[h!]
\resizebox{0.95\textwidth}{0.35\textheight}{
\setlength\arrayrulewidth{1.5pt}
\begin{tabular}{cccccccccccccccccccccccccc}
  \hline
 && \textbf{1} & \textbf{2} & \textbf{3} & \textbf{4} & \textbf{5} & \textbf{6} & \textbf{7} & \textbf{8} & \textbf{9} & \textbf{10} & \textbf{11} & \textbf{12} & \textbf{13} & \textbf{14} & \textbf{15} & \textbf{16} & \textbf{17} & \textbf{18} & \textbf{19} & \textbf{20} & \textbf{21} & \textbf{22} & \textbf{23} & \textbf{24} \\ 
 \multirow{16}{*}{\begin{sideways}\textbf{Week Base Future}\end{sideways}}
  &M3 lag 3 & \cellcolor[rgb]{1,1,0.692} 8.41 & \cellcolor[rgb]{1,1,0.755} 3.01 & \cellcolor[rgb]{1,1,0.528} 22.50 & \cellcolor[rgb]{1,0.778,0.27} 52.24 & \cellcolor[rgb]{1,1,0.536} 21.80 & \cellcolor[rgb]{1,1,0.786} 0.31 & \cellcolor[rgb]{1,1,0.788} 0.20 & \cellcolor[rgb]{1,1,0.718} 6.16 & \cellcolor[rgb]{1,1,0.727} 5.42 & \cellcolor[rgb]{1,1,0.768} 1.84 & \cellcolor[rgb]{1,1,0.787} 0.24 & \cellcolor[rgb]{1,1,0.785} 0.47 & \cellcolor[rgb]{1,1,0.789} 0.05 & \cellcolor[rgb]{1,1,0.79} 0.01 & \cellcolor[rgb]{1,1,0.79} 0.00 & \cellcolor[rgb]{1,1,0.788} 0.15 & \cellcolor[rgb]{1,1,0.79} 0.01 & \cellcolor[rgb]{1,1,0.774} 1.33 & \cellcolor[rgb]{1,1,0.781} 0.78 & \cellcolor[rgb]{1,1,0.726} 5.49 & \cellcolor[rgb]{1,1,0.719} 6.08 & \cellcolor[rgb]{1,1,0.677} 9.69 & \cellcolor[rgb]{1,0.771,0.265} 47.02 & \cellcolor[rgb]{1,1,0.747} 3.61 \\
\hline
\end{tabular}

}
\caption{Continuation of the Illustration of the percentage of the actually used regressors for each of the 24 hour time series of the day-ahead price. The regressors for the day-ahead price from lag 2 to 6 were dropped from the illustration.}
\label{tab_BBTRUE2}
\end{table}

In this regard Tables \ref{tab_BBTRUE} and \ref{tab_BBTRUE2} provide useful insight in the price process and decision making of market participants. The tables show the percentage on how often a given regressor was used by the algorithm for the final model throughout the time frame of the rolling estimation window. To provide a better overview, colors range from light yellow as 0 percent to red which represents 100 percent. This is shown in each column for each of the different 24 models for every hour. For instance, 0 percent means that this regressor was not used once in all the days of May 2016 up to April 2017 and had therefore too little additional explanatory power beyond the other regressors. \NEW{Due to the specific modeling setup, some of the regressors were not available throughout all 28 forecasting days. Therefore regressors were only counted if they had a chance to be chosen by the algorithm.} The first 24 lines correspond to the day-ahead price of each hour one day ago. At first sight it can be deducted, that the 24 hours of a day behave quite different from each other. For some hours, e.g. \NEW{8 to 11} \OLD{7 to 12}, their own lag of one was in most cases not even relevant, but for 
the evening hours from 18 to 24 they are highly relevant. However, it can be easily seen that the last hour of the day hour 24 is a very good regressor for almost every hour of the day, except the late afternoon and early evening hours. This finding is very similar to the one of \cite{ziel2016day}, which use a related scheme for their investigation of the relevance of different regressors. But comparing both analyses leads also to another finding. By including the future data into the dataset, the autoregressive structure seems to diminish. The majority of cells in the first 24 lines and columns of Table \ref{tab_BBTRUE} have a value of 0, meaning that this lag 1 regressor was not included even once. The following lags of the day-ahead prices had similar behavior but with diminishing percentages the higher the investigated lag was. The weekly lag, e.g. lag 7, tended to have some more relevant regressors than e.g. lag 6, but the influence was also much smaller than the lag 1 autoregressive regressors. Lag 2 to lag 6 was cut out of the table due to space issues.

A remarkable finding, which might explain the comparably little amount of autoregressive lags, can be seen by investigating the daily futures. These futures seem to have a vast impact on the electricity price, the best of them with maturity 3 lag 1 was chosen, if we combine the peak and base variant, for almost every hour every time. It is also no surprise that maturity 2 with lag 0, maturity 3 with lag 1, maturity 4 with lag 2 were that important, as due to the before mentioned data structure these are the day futures for the corresponding day-ahead price. This is also an explanation for the extremely good performance of the model for the first week, as these regressors are deterministically known when the day-ahead price is negotiated. \OLD{The maturity 1 day future seem to have almost no impact, as this corresponds to a day before the day-ahead price.} Another interesting finding can be deducted from the comparison of peak and base day futures. We can clearly see that peak products have a tremendous impact from hour 9 to \NEW{20}\OLD{19} and almost no impact on other hours. Some may think that this should be clear, as these products are by definition designed to concern the hours 9 to 20 according to EEX. However, the lasso-algorithm we have used did not know that a priori, every hour had been given the same set of possible regressors of peak products and the algorithm had to make a choice only based on the information inside the time-series itself to make a decision of whether to include the regressor or not. The same holds true for the base day products, where we can observe exactly the opposite result. Here the base products seem to have little influence on the peak hours, as these are captured by the day peak products. \NEW{A very interesting and novel finding can be deducted from the day base future products with maturity 3 lag 0, maturity 4 and lag 1 and so one for the evening hours 20 to 24. It seems that these products were always relevant, even though they actually are the future price for the day after the current electricity price. One likely explanation is that the evening hours of a day are indeed very close to the early morning hours of the next day for which this product is traded. Hence, investors might already take a look at the futures for the next day to make conclusions about the evening hours of the current electricity price.}

\OLD{For the week and weekend futures we notice moderate importance overall, some of the products seem to have no influence at all while the weekend base future with maturity 12 for instance seems to have a substantial influence. For the week peak and base future products, we do not have very clear information on the influence. Only the week base future with maturity 4 and lag 3, which corresponds to the price for the recent week traded four weeks ago seem to have a decisive impact. The peak products of the weekend seem to have only no to little influence.}
\NEW{For the week base and peak future we notice moderate importance overall, while the week peak future seems to have a higher impact than the base future. We observe the same behavior as for the day futures, where the week futures which corresponded to the current day-ahead electricity price, e.g. maturity 3 lag 0, have the highest influence on the price formation. Also, week peak futures seem to have almost no influence outside of the peak hours from 9 to 20.

The weekend futures however seem to have overall almost no impact, no matter if traded as base or peak product.} One possible reason for that could be that there were only \NEW{a few} \OLD{little} trades for that product available and the market for these niche products is still in an early stage.

\OLD{The month futures also had only moderate impact on some of the 24 time-series. The month base future was only chosen very rarely. The peak month future however seem to have at least some impact on the hours of the early morning and late evening.} \NEW{The month futures seem to account for a moderate amount of the variation in the price formation process. They gain their influence most likely throughout the last forecasting days, as they provide long-term stability for the estimation of the price formation. For short-term forecasting the lack of variation in month futures might be a reason why they may not contribute a lot to the estimation precision.} 

For the dummy variables from Monday to Sunday it can be obtained that especially the Monday and the combined Sunday plus holiday dummies are of \NEW{importance}\OLD{meaning}. However, \OLD{their influence} \NEW{the influence of all other weekday dummies} is heavily weakened due to the future products, as they capture the deterministic \OLD{seasonal} \NEW{recurring daily} structure as well. \OLD{It is interesting to notice that the Saturday dummy introducing the weekend had almost no influence on the estimation. This could be explained by the fact that we used weekend futures which might have had an equivalent but yet more detailed impact than the Saturday dummy.}

\NEW{The periodic B-spline regressors which represent the seasons of a year seem to have a noticeable impact on the price formation. Especially the winter regressor has had a vast impact on some of the peak hours. The spring regressor however exhibited the least impact out of all seasonal regressors.}

\NEW{The proposed future-model is also capable of producing realistic simulations for the day-ahead electricity price. By bootstrapping over the residual time series and iteratively forecasting the price of the hour of the next day for all 24 sub-models, we can produce random simulation patterns, which are suitable for modelling the whole probability density of the price series. To construct such a forecast we estimate the mean forecast for a specific hour as described beforehand but add a randomly chosen residual of the in-sample residuals. This is done for all 24 hours of the day. Afterwards we use this data point to forecast the data points of the next day by again forecasting the mean and adding a randomly chosen residual. This is repeated iteratively until we get a simulation path for all 28 days, e.g. 672 hours. Overall, we decided to repeat this procedure 1000 times to be able to conduct a probabilistic analysis of all simulation paths. 

\begin{figure}
\centering
\begin{subfigure}[b]{0.8\textwidth}
   \includegraphics[width=1\linewidth]{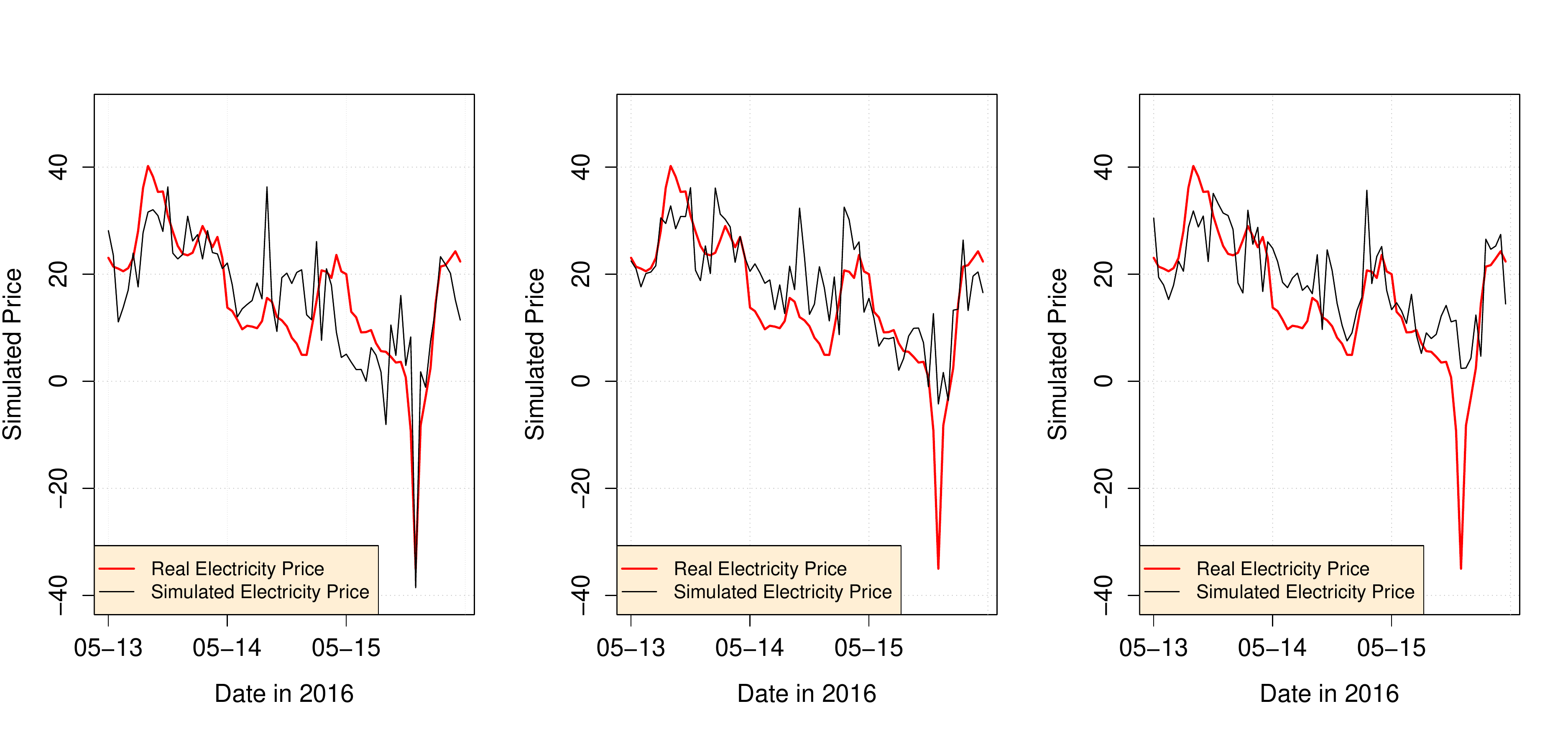}
   \caption{}
   \label{sim_days} 
\end{subfigure}

\begin{subfigure}[b]{0.8\textwidth}
   \includegraphics[width=1\linewidth]{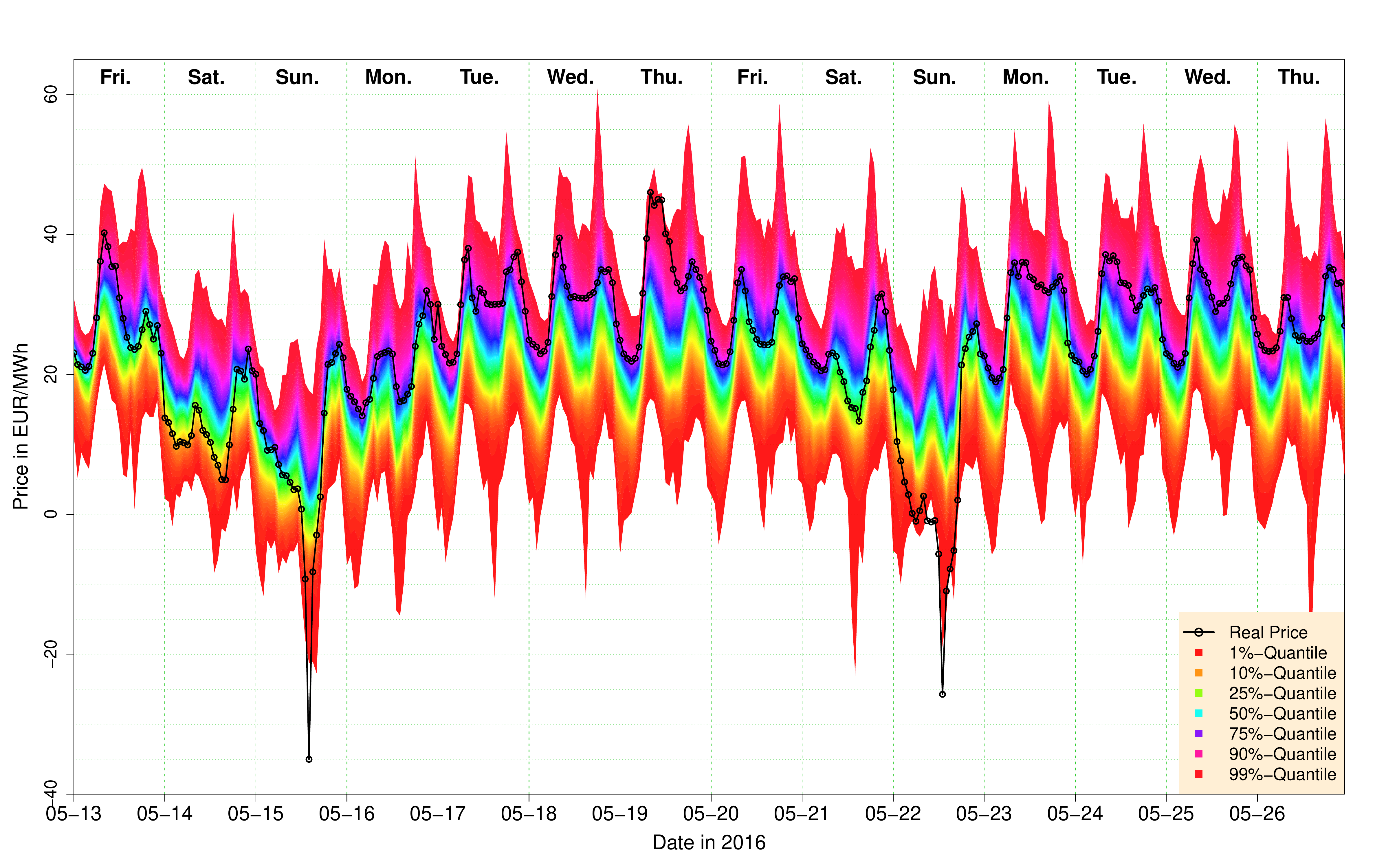}
   \caption{}
   \label{sim_quant}
\end{subfigure}

\caption{(a) Three selected simulated day-ahead electricity price patterns in comparison with the real day-ahead electricity price for 72 hours starting from the 13.05.2016. (b) Mean and Quantile Forecasts for the electricity price starting from the 13.05.2016 to 26.05.2016 in comparison to the real electricity price.}
\end{figure}

Figure \ref{sim_days} shows exemplarily three of these simulation paths. These three paths were chosen manually to show different features of these paths. The left hand picture shows the simulations path for three days including the 15.05.2016, where the electricity price reached negative prices up to -35.02 EUR/MWh. This is depicted as a red line in the picture. It is also shown that among the 1000 different samples there was at least one simulation path which was able to capture this behavior, as the severe negative movement during the afternoon of that day was captured quite well. However, this is not always the case, as the two other pictures show. These two simulation paths recognize the tendency of the falling price but are not able to detect the severe price spike at the 15.05.2016. But as these are just three samples out of 1000 different paths, it is not possible to make statistical valid statements about the overall quality of the simulation paths.

Therefore Figure \ref{sim_quant} shows the 99 percentiles of all simulation paths for the time frame of 13.05.2016 to 26.05.2016. We left out the remainder of $28-14=14$ days due to illustrative purposes. It can be easily obtained that on the 15.05.2016 as well as on the 22.05.2016 the electricity price, depicted as solid black line with dots, exhibited negative price spikes. Both price spikes laid close to the red quantile interval, but still outside the 99\% quantile. However, the negative prices close to the most severe price spikes usually still laid within the corresponding quantile, which indicates a that the area shows a realistic picture of the true density. The picture overall shows not only that our model is able to produce realistic simulation results, but also a realistic model for the whole probabilistic density of electricity prices. The density is heavily left-skewed towards negative prices during the midday hours, which is in accordance with the real life observations that during these hours electricity consumption usually declines and leads to the electricity price therefore being more prone to unexpected shocks in the electricity production, e.g. large amounts of solar and wind energy. It is also remarkable that these price events are observable in the simulation pattern for many days in the future. This also shows the importance of futures especially for short- to mid-term forecasting.
}

\section{Discussion and Conclusion}
Overall we have shown that it is possible and beneficial to utilize future products for modeling day-ahead electricity prices. The complex dependency structure and the difference between the two markets stood out as the two main challenges for this modeling approach. However, by employing a comprehensive model setup it was possible to get a first impression on how such a dependency structure can be modeled. The chosen approach has proven to increase forecasting precision over several other strong benchmark models in the short- as well as in the medium-term for up to four weeks. The deterministic future product components decreased uncertainty and therefore helped the model to tackle the issue of regressor uncertainty.

\NEW{We have shown that only incorporating future products of the current day as usually done in HPFC-approaches may not fully account for the information structure given in day-ahead prices. Table \ref{tab_BBTRUE} shows clearly that especially day base and day peak futures with higher lags but corresponding maturities, e.g. the day base future with maturity 3 with lag 1, do have a substantial impact on the price formation of day-ahead electricity prices. It could therefore be argued, that investors, who contracted such a future product, still remember the agreed settlement price and trade accordingly. If this is the case, constructing the HPFC by only using futures of the current day to make forecasts for the day-ahead prices may not be sufficient enough.}

\OLD{However, with an increasing forecasting horizon the deterministic structure diminishes, as e.g. day futures were only available for up to six days. But when they are included into the model and longer forecasting horizons are considered they have to be forecasted as well. This again leads to the described problem of stacking errors for the regressors. Figure 5 shows this drastically as starting with day six, the other benchmark models seem to outperform the Future-Model for a small amount of hours. The hours of week four often had hours, where the other benchmarks performed better. Comparing this to Figure 1 provides a deeper insight into this, as we can see that for the fourth week we usually have only four deterministic future values, the maturity 4 week future and the month futures as base and peak products. Moreover, for forecasting our regressors we only used AIC-selected AR, which do not incorporate interdependencies between all the time series. We therefore think that forecasting accuracy can still be improved by either including new deterministic regressors or by a more sophisticated approach for the forecasting of regressors.}

\NEW{By using a bootstrap based method we were able to simulate different electricity price scenarios with our model. Aggregating over all these simulation paths, we conducted a study of the probabilistic capability of our model. It was shown that futures can play an important role in forecasting the whole probability density of electricity prices. However, as our main purpose was to forecast the mean of electricity prices the study of the probabilistic density could be much more extended. Giving an accurate measure of evaluation for probabilistic events paired with a much larger forecasting horizon could provide much more insights into the probabilistic capability of our model approach.}

For our modeling we used a very specific regressor setup which was based on observation, testing and common sense. Nevertheless, one can think of other setups, where other products or other updating schemes could be used. For instance, we left out the off-peak month future or the daily changes in the month future values. The model setup could be also switched by creating and matching the hourly price forward curves to the different hours of the day. This could improve the model setup as this would allow us to match every hour with its specific future value rather than using e.g. the base value which actually accounted for every hour of that day.

Moreover, we only had the end-of-day data for future products available. But in reality, the trading decision for the day-ahead prices has to be made prior to 12:30 pm for which not only the end-of-day data of the previous day is available but also the intraday values of future products. The Future-Model could therefore be improved when intraday future data is included, as it is likely that these time-series contain updated and better information about the upcoming day-ahead prices.

Finally, it is still left for future research, if these future products are actually a better substitution for conventional regressors, like production and consumption. Modeling both regressors together however, could lead into a situation where e.g. future products are dropped out by the algorithm because renewable production can describe the day-ahead prices better in-sample. But this in turn leads to a problem that if renewables are almost impossible to forecast the forecasting accuracy of that model would still be inferior to our model. It should be therefore thoroughly analyzed in which scenarios it is indicated to use which set of regressors.

\bibliographystyle{apalike}
\bibliography{future_model}

\end{document}